\documentclass[10pt,a4paper,twocolumn]{IEEEtran}
\IEEEoverridecommandlockouts                              

\usepackage{amsmath}
\usepackage{color}
\usepackage{graphicx}
\usepackage{latexsym}

\usepackage{epsfig} 

\title{Minimal Inputs/Outputs for a Networked System}

\author{Tong Zhou
\thanks{This work was supported in part by the NNSFC under Grant 61573209 and 51361135705.}
\thanks{T.Zhou is with the Department of Automation and TNList, Tsinghua University, Beijing, 100084,
CHINA. (Tel: 86-10-62797430; Fax: 86-10-62786911; e-mail:
tzhou@mail.tsinghua.edu.cn.)}}

\begin{document}
\renewcommand{\thefootnote}{\fnsymbol{footnote}}
\maketitle 
\renewcommand{\thepage}{6--\arabic{page}}
\setcounter{page}{1}

\begin{abstract}
This paper investigates the minimal number of inputs/outputs required to guarantees the controllability/observability of a system, under the condition that its state transition matrix (STM) is prescribed. It has been proved that this minimal number is equal to the maximum geometric multiplicity of the system STM. The obtained conclusions are in sharp contrast to those established for the problems of finding the sparest input/output matrix under the restriction of system controllability/observabilty, which have been proved to be NP-hard, and even impossible to be approximated within a multiplicative factor. Moreover, a complete parametrization is also provided for the input/output matrix of a system with this minimal input/output number.

\end{abstract}

\IEEEpeerreviewmaketitle

\section{Introduction}

With the development of network and communication technologies, etc., various new challenging theoretical issues arise in the analysis and synthesis of a networked  system, which can be regarded as one special kind of large scale systems that have attracted extensive research interests since almost half a century ago \cite{siljak78,sbkkmpr11,zhou15,zz16}. Some examples are efficient utilization of the structure information like sparsity etc., in reducing computational costs in distributed estimation and control, analysis for their stability, controllability, observability \cite{emccb12,np13,pka16,ra13,zhou15}. Other issues include structure identification using measurement data, parameter estimation for a power law used in sparsity descriptions, etc. \cite{siljak78,sbkkmpr11,xz14}. Among these investigations, variable selections for insuring system controllability/observability are of extreme importance, as a brute force search is usually computationally prohibitive and controllability/observability is essential for a system to work satisfactorily \cite{emccb12,lsb11,pka16,scl15,zhou15}.

Recently, it has been discovered that under various well encountered situations, the problem of finding the minimal number of directly manipulatable system states for guaranteeing system controllability, as well as the problem of finding the minimal number of directly measurable system states for insuring system observability, which are respectively called as a minimal controllability problem (MCP) and a minimal observability problem (MOP) there, are NP-hard, and are also impossible to be approximated within a multiplicative factor \cite{pka16,scl15}. It has also been observed there that some simple heuristic methods usually give an "acceptable" small amount of system states that lead to a controllable/observable system.

These results are of great theoretical values, and have settled a long standing and important issue in system and control theories \cite{wj01}.

On the other hand, \cite{sm68,Sontag98,yzdwl13} investigated the minimal number of inputs for insuring system controllability. \cite{Sontag98} leaves it as an exercise to prove that in order to guarantee that a system is controllable, it is necessary that the number of its input is at least equal to the maximum geometric multiplicity of the system STM. In \cite{sm68}, in order to establish its conclusions, the Jordan canonical form is adopted in decomposing the state space of a system into controllable and uncontrollable subspaces. This is generally impossible, as a real square matrix may have complex eigenvalues and eigenvectors, but the state space of a system usually consists of only real vectors. In \cite{yzdwl13}, the Jordan canonical form is once again adopted and the sufficiency of its condition is demonstrated only through some numerical examples. Moreover, in their construction of the input matrix $B$, $B=PQ$ is utilized which in general can not guarantee that the constructed input matrix is real. Here, the matrix $P$ is nonsingular that transforms the STM $A$ into a Jordan canonical form, and the matrix $Q$ is constructed to have the smallest number of nonzero elements. These means that conclusions of \cite{sm68,yzdwl13} are only valid when a system input matrix is permitted to take complex values, which is often not possible.

In this paper, we reinvestigate this minimal input determination problem under the constraint that the input matrix is real. It has been made clear that contrary to the MCPs and MOPs discussed in \cite{pka16,scl15}, these problems can be clearly solved using an analytic form depending only on the system STM. It has been proved that in order to guarantee the controllability of a linear time invariant (LTI) system, the minimum number of inputs is simply equal to the maximum geometric multiplicity of the system STM, which can be calculated in principle. Moreover, the minimum number of outputs for system observability is also equal to the maximum geometric multiplicity of the system STM. These conclusions suggest that to reduce the number of inputs and/or outputs without sacrificing system controllability and/or observability, it is preferable to design a system with its STM having distinct eigenvalues. In addition, a complete parametrization is provided for the input/output matrices of a controllable/observable system with the minimal input/output number.

The rest of this paper is organized as follows. In the next section, we give a description for the minimal control selection problem (MISP) and the minimal output selection problem (MOSP), and present some preliminary results. The main results are reported in Section III, which at first gives a clear answer to the MISP, and then extends the results to the MOSP using the dualities between controllability and observability of a system. Some special situations are discussed in Section IV which serves as numerical examples to illustrate the theoretical results of Section III. Section V discusses parametrization of the input/output matrices for a controllable/observable system with the minimal number of inputs/outputs. Finally, Section VI concludes this paper.

The following notation and symbols are adopted in this paper. ${\cal R}^{m\times n}$ and ${\cal C}^{m\times n}$ are utilized to represent the $m\times n$ dimensional real and complex linear spaces. When $m$ and/or $n$ are equal to $1$, they are usually omitted. ${\rm\bf Span}\!\{x_{i}|_{i=1}^{n}\}$ stands for the space consisting of all the linear combinations of the vectors $x_{1}$, $x_{2}$, $\cdots$, $x_{n}$, while ${\rm\bf col}\!\{x_{i}|_{i=1}^{L}\}$ the vector/matrix stacked by $x_{i}|_{i=1}^{L}$ with its $i$-th row block vector/matrix being $x_{i}$, and ${\rm\bf diag}\!\{x_{i}|_{i=1}^{n}\}$ a diagonal matrix with its $i$-th diagonal element being $x_{i}$. $0_{m}$ and $0_{m\times n}$ stand respectively for the $m$ dimensional column vector and the $m\times n$ dimensional matrix with all elements being zero. The superscript $T$ and $H$ are used to denote respectively the transpose and the conjugate transpose of a matrix/vector, while $\bar{\cdot}$ the conjugate of a complex number/variable/vector/matrix.

\section{Problem Formulation and Preliminary Results}

Consider a continuous LTI dynamic system $\bf\Sigma$ with its state space model being
\begin{equation}
{\bf\Sigma}: \hspace{0.1cm} \frac{{\rm d}x(t)}{{\rm d} t}=Ax(t)+Bu(t),\hspace{0.15cm}
y(t)=Cx(t)+Du(t)   \label{eqn:1}
\end{equation}
in which $x(t)$ is the system state vector which is $n$ dimensional and real valued, while $A$, $B$, $C$ and $D$ respectively a $n\times n$, $n\times q$, $p\times n$ and $p\times q$ dimensional real valued matrices.

In system designs, an interesting problem is to find the minimum number of inputs, which is equivalent to the smallest $q$, such that there exists a matrix $B\in {\cal R}^{n\times q}$ that makes the corresponding LTI system ${\rm \bf \Sigma}$  controllable \cite{wj01,zhou15}. To clarify differences between this problem and those discussed in \cite{pka16,scl15} in which the sparest matrix $B$ is searched under the condition that the value of $q$ is fixed, we call it the minimal input selection problem (MISP) in this paper. In practical applications, it is also often interesting to know that in order to construct an observable system, how many outputs are required \cite{wj01,zhou15}. More precisely, with respect to the above continuous LTI system $\bf\Sigma$, what is the minimal dimension of the output vector $y(t)$, under the restriction that there exist an output matrix $C$ and a direct coupling matrix $D$, such that the system $\bf\Sigma$ is observable.
This problem is called a minimal output selection problem (MOSP) in this paper.

When an LTI system with its input-output relation being described by Equation (\ref{eqn:1}) is controllable/observable, we sometimes also use a short expression in this paper for brevity, which states as that the matrix pair $(A,\;B)$/$(A,\;C)$ is controllable/observable.

To investigate this MISP, we at first need the following results on system controllability/observability verifications.

\hspace*{-0.45cm}{\bf Lemma 1.} A system with its input-output relation being described by Equation (\ref{eqn:1}) is controllable, if and only if for every complex scalar $\lambda$ and every nonzero complex vector $x$ satisfying $x^{H}A=\lambda x^{H}$, $x^{H}B\neq 0$. Or equivalently, the controllability matrix $[B\;\; AB\;\; A^{2}B\;\; \cdots\;\; A^{n-1}B]$ has a full row rank (FRR). In addition, the system $\bf\Sigma$ is observable, if and only if for every complex scalar $\lambda$ and every nonzero complex vector $y$ satisfying $Ay=\lambda y$, $Cy\neq 0$.

These conditions are widely known in linear system theory, which are extensively called as the PBH test or the rank test \cite{zdg,zhou15}.

\section{Main Results}

For a concise presentation, it is assumed throughout this paper, without any loss of generality, that the $n\times n$ dimensional real matrix $A$ has $k_{r}$ distinct real eigenvalues $\lambda_{r,i}$, $i=1,2,\cdots,k_{r}$, and $k_{c}$ distinct complex eigenvalues $\lambda_{c,i}$, $i=1,2,\cdots,k_{c}$. Moreover, let $x_{*,i}(j)$, $j=1,2,\cdots,p_{*}(i)$, $*=r,c$, denote a set of the left eigenvectors of this matrix associated with the same eigenvalue $\lambda_{*,i}$, which are linearly independent and spans the null space of the matrix $\bar{\lambda}_{*,i}I_{n}-A^{T}$. Furthermore, let $p_{max}$ denote the maximum value among the numbers $p_{*}(i)$, $i=1,\; 2,\;\cdots,\; k_{*}$; $*=r$, $c$. That is,
\begin{equation}
p_{max} = \max\left\{\max_{1\leq i\leq k_{r}}p_{r}(i),\; \max_{1\leq i\leq k_{c}}p_{c}(i)\right\}
\end{equation}

Note that the matrix $A$ is real. It can be straightforwardly proved that any left eigenvector of the matrix $A$ associated with an eigenvalue $\lambda_{*,i}$ is in fact a right eigenvector of the matrix $A^{T}$ associated with its eigenvalue $\bar{\lambda}_{*,i}$. This means that the vectors $\left.x_{*,i}(j)\right|_{j=1}^{p_{*}(i)}$ are well defined,  and $p_{*}(i)\geq 1$. In addition, for each $*=r,c$ and $i=1,2,\cdots, k_{*}$, $p_{*}(i)$ is in fact the geometric multiplicity of the matrix $A$ associated with its eigenvalue $\lambda_{*,i}$\cite{hj91}.

On the basis of Lemma 1, the following conclusion is established, which gives a necessary and sufficient condition on an input matrix $B$, such that the corresponding system $\rm\bf\Sigma$ is controllable.

\hspace*{-0.45cm}{\bf Lemma 2.}  For each $*=r,c$ and each $i=1,2,\cdots,k_{*}$, define matrix $X_{*,i}$ as
\begin{equation}
X_{*,i}=\left[\;x_{*,i}(1)\;\;  x_{*,i}(2)\;\; \cdots \;\; x_{*,i}(p_{*}(i))\;\right]
\label{eqn:2}
\end{equation}
Then, a system with its state transitions being described by Equation (\ref{eqn:1}) is controllable, if and only if the matrix $B^{T}X_{*,i}$ is of full column rank (FCR) with every $*=r,c$ and every  $i=1,2,\cdots,k_{*}$.

\hspace*{-0.45cm}{\bf\it Proof}: Let the set ${\cal X}$ represent ${\cal C}$ when $*=c$ and ${\cal R}$ when $*=r$. From the definitions of the matrix $X_{*,i}$ and properties of the eigenvectors of a matrix, it is clear that for each $*=r,c$ and each $i=1,2,\cdots,k_{*}$, $x$ is a left eigenvector of the matrix $A$ associated with its eigenvalue $\lambda_{*,i}$, if and only if there exists a nonzero vector $\alpha\in{\cal X}^{p_{*}(i)}$, such that $x=X_{*,i}\alpha$. The proof can now be completed through a direct application of the PBH test in Lemma 1. This completes the proof. \hspace{\fill}$\Diamond$

It is worthwhile to mention that a left eigenvector of a real valued square matrix may still be complex valued \cite{hj91}. On the other hand, for a practically realizable system, its input matrix is usually required to be real valued. This real-complex mixture asks careful investigations about the MISP. On the other hand, note that when the matrix $A$ is real valued, it is well known that if a vector $x\in {\cal C}^{n}$ is a left eigenvector of this matrix associated with a complex eigenvalue $\lambda$, then, the complex number $\bar{\lambda}$ is also one of its eigenvalues, and the vector $\bar{x}$ is a left eigenvector associated with this eigenvalue $\bar{\lambda}$ \cite{hj91}. It can therefore be declared that if $k_{c}\neq 0$, then, it is certainly an even number. Hence, it can be assumed, without any loss of generality, that $\lambda_{i+k_{c}/2}=\bar{\lambda}_{i}$ and $X_{i+k_{c}/2}=\bar{X_{i}}$, $i=1,2,\cdots,\frac{k_{c}}{2}$, provided that $k_{c}\geq 1$. This assumption is adopted throughout the rest of this paper.

In order to give a clear description on the minimal $q$, the STM $A$ is at first expressed through its Jordan canonical form. From the assumption that the eigenvalues $\lambda_{*,i}$ are different and each of them has $p_{*}(i)$ independent left eigenvectors, $i=1,2,\cdots,k_{*}$, $*=r$ or $c$, it can be declared from results on matrix analysis \cite{hj91} that, associated with each eigenvalue $\lambda_{*,i}$, there are $p_{*}(i)$ Jordan blocks. Denote these Jordan blocks by $J_{*,i,j}$, and assume their dimensions being $m_{*,i,j}$ respectively, $j=1,2,\cdots, p_{*}(i)$. Obviously, $m_{*,i,j}$ is a positive integer not smaller than $1$. Moreover, for each Jordan block, there exists an $n\times m_{*,i,j}$ dimensional matrix $T_{*,i,j}$ which is of FCR and satisfies
\begin{equation}
T_{*,i,j}J_{*,i,j}=AJ_{*,i,j}
\label{eqn:13}
\end{equation}
and the vector set consisting of the first columns of the matrices $T_{*,i,j}|_{j=1}^{p_{r}(i)}$ are linearly independent. Furthermore, the matrix $T_{*,i,j}$ is real when the associated eigenvalue is real, and is generally complex  when the associated eigenvalue is complex.

For a given scalar $\alpha_{*,i,j}$, define a matrix $\hat{B}_{*,i}$ as
\begin{displaymath}
\hat{B}_{*,i}\!=\!\left[\!{\rm\bf diag}\!\left\{\!\left[\!\begin{array}{c} 0_{m_{*,i,j}\!-\!1} \\ \alpha_{*,i,j}\end{array}\!\right]_{j=1}^{p_{*}(i)}\!\right\},\;
0_{\sum_{j=1}^{p_{*}(i)}m_{*,i,j}\times (p_{max}\!-\!p_{*}(i))} \!\right]
\end{displaymath}
in which $\alpha_{*,i,j}$ belongs to the set $\cal R$ when $*=r$, and belongs to the set $\cal C$ when $*=c$. On the basis of these matrices, construct an input matrix $B$ as
\begin{eqnarray}
B &=& \sum_{i=1}^{k_{r}}\left[T_{r,i,j}\; T_{r,i,2}\; \cdots\; T_{r,i,p_{r}(i)}\right]\hat{B}_{r,i}+\nonumber\\
& & 2\sum_{i=1}^{k_{c}/2}\Re\left\{\left[T_{c,i,j}\; T_{c,i,2}\; \cdots\; T_{c,i,p_{r}(i)}\right]\hat{B}_{c,i}\right\}
\label{eqn:14}
\end{eqnarray}
Here, $\Re\left\{\cdot\right\}$ denote the operation of taking the real part of a matrix. Using this particularly constructed input matrix, the following results are obtained, which reveal the minimal input number for System $\rm\bf\Sigma$ being controllable.

\hspace*{-0.45cm}{\bf Theorem 1.} There exists a $n\times q$ dimensional real matrix $B$, such that a system with its state transitions being described by Equation (\ref{eqn:1}) is controllable, if and only if $q$ is not smaller than $p_{max}$.

\hspace*{-0.45cm}{\bf\it Proof}: Assume that there exists a matrix $B\in {\cal R}^{n\times q}$ with $q<p_{max}$, such that the matrix pair $(A,B)$ is controllable. Let ${\cal I}$ denote the set consisting of the indices of the eigenvalues of the matrix $A$, such that the maximum number of the associated linearly independent left eigenvectors achieves $p_{max}$. That is,
\begin{equation}
{\cal I}={\cal I}_{r}\bigcup {\cal I}_{c}
\end{equation}
in which
\begin{eqnarray*}
& & {\cal I}_{r}=\left\{\;i \left|\; p_{r}(i)=p_{max}, \;\; 1\leq i\leq k_{r}\right.\right\} \\
& & {\cal I}_{c}=\left\{\;i \left|\; p_{c}(i)=p_{max}, \;\; 1\leq i\leq k_{c}\right.\right\}
\end{eqnarray*}

From the definitions, it is clear that both the set ${\cal I}_{r}$ and the set ${\cal I}_{c}$ might be empty, but it is certain that they can not be simultaneously empty.

Assume that the set ${\cal I}_{c}$ is not empty. For an arbitrary positive integer $i\in {\cal I}_{c}$, from the definition of the matrix $X_{c,i}$ given by Equation (\ref{eqn:2}), we have that the dimension of the matrix $B^{T}X_{c,i}$ is $q\times p_{max}$ , which can not be FCR when $q<p_{max}$. This contradicts with Lemma 2. Similar arguments apply when the set ${\cal I}_{r}$ is not empty. Hence, in order to guarantee the controllability of the matrix pair $(A,\;B)$, the matrix $B$ must have at least $p_{max}$ columns.

On the other hand, from Equation (\ref{eqn:13}) and the fact that the STM $A$ is real, as well as the arrangements of the eigenvalues of the STM $A$,
it is obvious that $\bar{T}_{*,i,j}\bar{J}_{*,i,j}=A\bar{J}_{*,i,j}$, which further implies that $\bar{T}_{c,i,j}{J}_{c,i+k_{c}/2,j}=A{J}_{*,i+k_{c}/2,j}$ is valid for each $i=1,2,\cdots,\frac{k_{c}}{2}$ and $j=1,2,\cdots,p_{c}(i)$. Define a matrix $T$ as
\begin{equation}
T=[T_{r}\;\; T_{c}]
\label{eqn:3}
\end{equation}
with
\begin{eqnarray*}
& & T_{r}\!=\!\left[T_{r,i,j}\right]_{j=1,i=1}^{j=p_{r}(i),i=k_{r}} \\
& & T_{c}\!=\!\left[\left[T_{c,i,j}\right]_{j=1,i=1}^{j=p_{c}(i),i=k_{c}/2}\;  \left[\bar{T}_{c,i,j}\right]_{j=1,i=1}^{j=p_{c}(i),i=k_{c}/2}\;\right]
\end{eqnarray*}
From the result that eigenvectors of a matrix associated with different eigenvalues are linear independent \cite{hj91} and the assumption that the first columns of the matrices $T_{*,i,j}|_{j=1}^{p_{r}(i)}$ are linearly independent, it can be straightforwardly proved that the matrix $T$ is invertible. Moreover,
\begin{eqnarray}
& & T^{-1}AT={\rm\bf diag}\left\{\left.J_{*,i,j}\right|_{j=1,i=1,*=r}^{j=p_{*}(i),i=k_{*},*=c}\right\}
\label{eqn:15}\\
& & T^{-1}B={\rm\bf col}\left\{\left.\hat{B}_{r,i}\right|_{i=1}^{k_{r}},\; \left.\hat{B}_{c,i}\right|_{i=1}^{k_{c}/2},\; \left.\bar{\hat{B}}_{c,i}\right|_{i=1}^{k_{c}/2}\right\}
\label{eqn:16}
\end{eqnarray}
Note that for an arbitrary complex number $\lambda\in{\cal C}$,
\begin{equation}
\left[\lambda I_{n}\!-\!A, \; B\right]\!=\!T\!\left[\lambda I_{n}\!-\!T^{-1}AT, \; T^{-1}B\right]\!{\rm\bf diag}\{T^{-1},\;I_{p_{max}}\}
\label{eqn:17}
\end{equation}
It is clear that the matrix $\left[\lambda I_{n}-A, \; B\right]$ is always of FRR, if and only if each $\alpha_{*,i,j}$ in the definition of the matrix $B$ is not equal to zero. Hence, through selecting an appropriate value for $\alpha_{*,i,j}$, an input matrix $B$ can be constructed which has exactly $p_{max}$ columns, and the associated matrix pair $(A,\;B)$ is controllable.

This completes the proof. \hspace{\fill}$\Diamond$

Theorem 1 makes it clear that in order to construct a controllable system, the minimal number of inputs is exactly equal to the maximum geometric multiplicity of the STM. The same results have also been given in \cite{sm68,Sontag98,yzdwl13}. In \cite{Sontag98}, however, it is only declared that $q\geq p_{max}$ is a necessary condition for the system $\rm\bf\Sigma$ to be controllable, and its proof is left as an exercise. On the other hand, the Jordan form of the STM $A$ is used in \cite{sm68} in decomposing the system state space into controllable and uncontrollable subspaces which is generally impossible, noting that system states usually take real values that can not be guaranteed by this decomposition in general. Moreover, \cite{yzdwl13} only illustrated sufficiency of the condition through a numerical  example, and in this illustration, the Jordan form was straightforwardly used once again, which can not guarantee that the constructed input matrix is real.

Now, consider the problem of finding the minimal number of outputs such that the system is observable.

Note that all the system matrices $A$, $B$, $C$ and $D$ are real valued. From Lemmas 1, it is clear that the observability of the matrix pair $(A,\;C)$ is equivalent to the controllability of the matrix pair $(A^{T},\;B^{T})$, which is well known in systems and control theory as the duality between system observability and system controllability \cite{zdg,zhou15}. In addition, observability of a system is not related to its direct coupling matrix $D$. These mean that the results of Section III can be directly applied to solve the above MOSP.

\hspace*{-0.45cm}{\bf Corollary 1.} There exists a matrix $C$ such that the system $\bf\Sigma$ is observable, if and only if the dimension of the output vector $y(t)$ is not smaller than the maximum geometric multiplicity of the STM $A$.

\hspace*{-0.45cm}{\rm\it Proof}: From the definitions of the left eigenvector and the right eigenvector of a matrix, it is obvious that a left eigenvector of the matrix $A^{T}$ is also a right eigenvector of the matrix $A$, and vice versa. The results can be immediately obtained from Theorem 1 through a utilization of the duality between the controllability and the observability of a system.

This completes the proof. \hspace{\fill}$\Diamond$

It is worthwhile to mention that while only continuous time systems are discussed in this paper, the results are also applicable to a discrete time system, noting that the PBH tests remain the same for these two kinds of LTI dynamic systems.

\section{Some Examples}
To illustrate the engineering significance of Theorem 1, we consider some simple and special but interesting situations, in which the STM $A$ has $n$ real eigenvalues $\lambda_{i}$, $i=1,2,\cdots, n$, and $n$ linearly independent left eigenvectors $x(i)$. Then, according to \cite{hj91}$, x(i)$ is certainly real valued for each $i\in \{1,2,\cdots,n\}$, and
\begin{equation}
A=T^{-1}\Lambda T,\hspace{0.5cm}  \Lambda={\rm\bf diag}\!\{\lambda_{i}|_{i=1}^{n}\},\hspace{0.5cm}   T={\rm\bf col}\!\{x^{T}(i)|_{i=1}^{n}\}
\label{eqn:8}
\end{equation}
Moreover, the inverse of the matrix $T$ always exists. For an arbitrary $\hat{B}\in{\cal R}^{n}$, define the system input matrix $B$ as $B=T^{-1}\hat{B}$. Then, this matrix is also certainly real valued.

On the other hand, denote the $i$-th row element of the vector $\hat{B}$ by $\hat{b}_{i}$, $1\leq i\leq n$. Direct algebraic manipulations show that
\begin{eqnarray}
& &\hspace*{-0.8cm} [B\;\; AB\;\; A^{2}B\;\; \cdots\;\; A^{n-1}B] \nonumber\\
& &\hspace*{-1.2cm}= T^{-1}[\hat{B}\;\; \Lambda\hat{B}\;\; \Lambda^{2}\hat{B}\;\; \cdots\;\; \Lambda^{n-1}\hat{B}] \nonumber\\
& &\hspace*{-1.2cm}= T^{-1}\left[\begin{array}{ccccc} \hat{b}_{1} & \lambda_{1}\hat{b}_{1} &  \lambda_{1}^{2}\hat{b}_{1} & \cdots & \lambda_{1}^{n-1}\hat{b}_{1}  \\
\hat{b}_{2} & \lambda_{2}\hat{b}_{2} &  \lambda_{2}^{2}\hat{b}_{2} & \cdots & \lambda_{2}^{n-1}\hat{b}_{2}  \\
\vdots & \vdots & \vdots & \ddots & \vdots \\
\hat{b}_{n} & \lambda_{n}\hat{b}_{n} &  \lambda_{n}^{2}\hat{b}_{n} & \cdots & \lambda_{n}^{n-1}\hat{b}_{n}  \end{array}\right] \nonumber \\
& &\hspace*{-1.2cm}= T^{-1}{\rm\bf diag}\!\{\hat{b}_{i}|_{i=1}^{n}\}\left[\begin{array}{ccccc} 1 & \lambda_{1} &  \lambda_{1}^{2} & \cdots & \lambda_{1}^{n-1} \\
1 & \lambda_{2} &  \lambda_{2}^{2} & \cdots & \lambda_{2}^{n-1}  \\
\vdots & \vdots & \vdots & \ddots & \vdots \\
1 & \lambda_{n} &  \lambda_{n}^{2} & \cdots & \lambda_{n}^{n-1}  \end{array}\right]
\label{eqn:9}
\end{eqnarray}
Note that the last matrix in the last line is a Vandermonde matrix, and its determinant can be analytically expressed using $\lambda_{i}|_{i=1}^{n}$ \cite{hj91}. Based on these results, the following conclusions are achieved.
\begin{eqnarray}
& & {\rm\bf det}\left([B\;\; AB\;\; A^{2}B\;\; \cdots\;\; A^{n-1}B]\right) \nonumber\\
&=& {\rm\bf det}^{-1}(T)\times\prod_{i=1}^{n}\hat{b}_{i}\times\prod_{i=1}^{n}\prod_{j=i+1}^{n}(\lambda_{i}-\lambda_{j})
\label{eqn:10}
\end{eqnarray}

Obviously, if all the eigenvalues of the STM $A$ are distinct, then, for an arbitrary input matrix $B$ with $\hat{b}_{i}\neq 0$ for each $i\in \{1,2,\cdots,n\}$, we have that ${\rm\bf det}\left([B\;\; AB\;\; A^{2}B\;\; \cdots\;\; A^{n-1}B]\right)\neq 0$. That is, the system controllability matrix is always of FRR, and the associated system is therefore always controllable according to Lemma 1. This means that under this situation, the minimal number of inputs for insuring system controllability is $1$.

However, if the STM $A$ has two or more than two eigenvalues that are equal to each other and there is just one input in the associated system, then, it can be declared from Equation (\ref{eqn:10}) that this system can not be controllable, no matter how the input matrix $B$ is selected. More precisely, under such a situation, Equation (\ref{eqn:10}) reveals that the determinant of the associated controllability matrix is constantly equal to $0$, which means that it can not have a FRR through {\it only} adjusting the element value of the input matrix $B$. Therefore, a controllable system can not be constructed.

Assume now that the STM $A$ just has two repeated eigenvalues. Then, Theorem 1 tells that two inputs can lead to a controllable system. In order to confirm this conclusion, assume without any loss of generality that $\lambda_{1}=\lambda_{2}$, and all the other eigenvalues of the STM $A$ are distinct and do not equal to $\lambda_{1}$. Construct an input matrix $B$ as
\begin{displaymath}
B=T^{-1} \left[\begin{array}{cc} {b}_{1} & 0 \\  0_{n-1} & {\rm\bf col}\!\{{b}_{i}|_{i=2}^{n}\} \end{array}\right]
\end{displaymath}
in which ${b}_{i}\neq 0$ for each $i=1,2,\cdots,n$. Note that when $i,j\in\{3,\;4,\;\cdots,\;n\}$ and $i\neq j$, we have from the adopted assumptions that $\lambda_{i}\neq \lambda_{1}$ and $\lambda_{i}\neq \lambda_{j}$. It can be directly proved that for each left eigenvector of the STM $A$, say $x$, there always exists a nonzero $n$ dimensional real valued vector $\alpha={\rm\bf col}\!\{{\alpha}_{i}|_{i=1}^{n}\}$, such that
\begin{displaymath}
x=\alpha_{1}x_{1}+\alpha_{2}x_{2}, \hspace{0.5cm}{\rm or} \hspace{0.5cm} x=\alpha_{i}x_{i}, \;\; 3\leq i\leq n
\end{displaymath}
Let $e_{i}$, $i=1,2,\cdots,n$, stands for the $i$-th canonical basis vector of the Euclidean space ${\cal R}^{n}$.
Then, from $TT^{-1}=I_{n}$ and $ T={\rm\bf col}\!\{x^{T}(i)|_{i=1}^{n}\}$, we further have that if this eigenvector is associated with the eigenvalue $\lambda_{1}$, which is equal to $\lambda_{2}$, then, ${\rm\bf col}\!\{{\alpha}_{i}|_{i=1}^{2}\}\neq 0$ and
\begin{eqnarray}
x^{T}B &=& [\alpha_{1}\;\;\alpha_{2}]\left[\begin{array}{c} x_{1}^{T} \\ x_{2}^{T}\end{array}\right]T^{-1} \left[\begin{array}{cc} {b}_{1} & 0 \\  0_{n-1} & {\rm\bf col}\!\{{b}_{i}|_{i=2}^{n}\} \end{array}\right] \nonumber\\
&=& [\alpha_{1}\;\;\alpha_{2}]\left[\begin{array}{c} e_{1}^{T} \\ e_{2}^{T}\end{array}\right] \left[\begin{array}{cc} {b}_{1} & 0 \\  0_{n-1} & {\rm\bf col}\!\{{b}_{i}|_{i=2}^{n}\} \end{array}\right] \nonumber\\
&=& [\alpha_{1}\;\;\alpha_{2}]\left[\begin{array}{cc} {b}_{1} & 0 \\ 0 & {b}_{2} \end{array}\right] \nonumber\\
&=& [\alpha_{1}b_{1}\;\;\alpha_{2}b_{2}]
\label{eqn:11}
\end{eqnarray}
Moreover, if this eigenvector is associated with the eigenvalue $\lambda_{i}$ with $i\in\{3,\;4,\;\cdots,\; n\}$, then, $\alpha_{i}\neq 0$ and
\begin{eqnarray}
x^{T}B&=&\alpha_{i}x_{i}^{T}T^{-1} \left[\begin{array}{cc} {b}_{1} & 0 \\  0_{n-1} & {\rm\bf col}\!\{{b}_{i}|_{i=2}^{n}\} \end{array}\right]\nonumber\\
&=&\alpha_{i}e_{i}^{T}\left[\begin{array}{cc} {b}_{1} & 0 \\  0_{n-1} & {\rm\bf col}\!\{{b}_{i}|_{i=2}^{n}\} \end{array}\right]\nonumber\\
&=&[0\;\; \alpha_{i}b_{i}]
\label{eqn:12}
\end{eqnarray}
From the assumption that $b_{i}\neq 0$ for each $i=1,2,\cdots,n$, it can now be declared that $x^{T}B\neq 0$ for every left eigenvector of the STM $A$. Hence, according to Lemma 1, the matrix pair $(A,\;B)$ is controllable. That is, two inputs are sufficient to lead to a controllable system.

These conclusions agree well with Theorem 1.

It is interesting to note here that under the aforementioned situation, the input matrix $B$ that makes the matrix pair $(A,\;B)$ controllable is not unique. For example,
\begin{displaymath}
B=T^{-1} \left[\begin{array}{cc} {b}_{1} & 0 \\  0 & b_{2} \\ {\rm\bf col}\!\{{b}_{i}|_{i=3}^{n}\} & 0_{n-2}\end{array}\right]
\end{displaymath}
with $b_{i}\neq 0$, $i=1,2,\cdots,n$, also leads to a controllable system.

The non-uniqueness of the input matrix leaves it a space for meeting other requirements, such as its sparseness, average control energy, etc. In fact, when the parameter $q$ of system matrices is fixed, many other interesting problems have been formulated and investigated for input matrix selections, see for example \cite{pka16,wj01} and the references therein.

\section{A Parametrization for All Desirable Input/Output Matrices}

In Section III, a necessary and sufficient condition is given for the existence of an input/output matrix $B$/$C$, such that the dynamic system $\rm\bf\Sigma$ is controllable/observable. In many engineering problems, there usually exist some other requirements on a system input/output matrix. For example, constraints on input energy, restrictions on the number of the states that can be directly affected/measured, etc. \cite{pka16,scl15,zhou15}. To satisfy these requirements, it appears desirable to have a parametrization for all system input/output matrices. In this section, a complete parametrization is at first given for all the input matrices $B$ which have the minimal column number and construct a controllable system with the STM $A$. Then, similar results are given through duality for system output matrices $C$ that have a minimal row number and construct an observable system with the same STM $A$.

To get this parametrization, for each $*=r,c$ and $i=1,2,\cdots,k_{*}$, define an integer $m_{*,i}$ as
\begin{equation}
m_{*,i}=\sum_{j=1}^{p_{*}(i)}m_{*,i,j}
\end{equation}
Moreover, for an arbitrary function of an integer variable $j$, define $\sum_{j=a}^{b}f(j)$ as $\sum_{j=a}^{b}f(j)=0$ whenever $b<a$. Then, we have the following results.

\hspace*{-0.45cm}{\bf Theorem 2.} The matrix pair $(A,\;B)$ is controllable with the matrix B having the minimal number of columns, if and only if there exist $m_{*,i}\times p_{max}$ dimensional matrices $\hat{B}_{*.i}$, $*=r,c$, $i=1,2,\cdots,k_{*}$, such that
\begin{equation}
B=T{\rm\bf col}\left\{ {\rm\bf col}\left\{\hat{B}_{r,i}|_{i=1}^{k_{r}}\right\},\;\;
{\rm\bf col}\left\{\hat{B}_{c,i}|_{i=1}^{k_{c}}\right\}\right\}
\label{eqn:4}
\end{equation}
is real, and the matrix
\begin{eqnarray}
\tilde{B}_{*,i}\!\!&=&\!\!\left[\begin{array}{ccc}
\hat{b}_{*,i}(1,1) & \hat{b}_{*,i}(m_{*,i,1}+1,1) & \cdots  \\
\hat{b}_{*,i}(1,2) & \hat{b}_{*,i}(m_{*,i,1}+1,2) & \cdots \\
\vdots & \vdots & \ddots  \\
\hat{b}_{*,i}(1,p_{*}(i)) & \hat{b}_{*,i}(m_{*,i,1}+1,p_{*}(i)) & \cdots  \end{array}\right. \nonumber\\
& &
\left.\begin{array}{c}
\hat{b}_{*,i}\left(\sum_{j=1}^{p_{*}(i)-1}m_{*,i,j}+1,1\right) \\
\hat{b}_{*,i}\left(\sum_{j=1}^{p_{*}(i)-1}m_{*,i,j}+1,2\right)  \\
\vdots \\
\hat{b}_{*,i}\left(\sum_{j=1}^{p_{*}(i)-1}m_{*,i,j}+1,p_{*}(i) \right)
\end{array}\right]
\label{eqn:5}
\end{eqnarray}
is of FCR for each $*=r,c$ and $i=1,2,\cdots,k_{*}$.

\hspace*{-0.45cm}{\rm\it Proof:} For each $*=r,c$ and $i=1,2,\cdots,k_{*}$, define a matrix $Z_{*,i}$ as
\begin{equation}
Z_{*,i}={\rm\bf diag}\left\{\left.\left[\begin{array}{c} 1 \\ 0_{m_{*,i,j}-1} \end{array}\right]\right|_{j=1}^{p_{*}(i)}\right\}
\label{eqn:6}
\end{equation}
From Equation (\ref{eqn:15}), it can be straightforwardly proved that, the matrix $X_{*,i}$ of Equation (\ref{eqn:2}) can be represented as
\begin{equation}
X_{*,i}=T^{-H}\left[\begin{array}{c} 0_{M_{*,i}-1}\times p_{*}(i) \\ Z_{*,i} \\
0_{(n-M_{*,i})\times p_{*}(i)} \end{array}\right]
\label{eqn:7}
\end{equation}
in which
\begin{displaymath}
M_{*,i}=\left\{\begin{array}{ll}
\sum_{j=1}^{i-1}m_{r,j} & *=r   \\
M_{r,k_{r}}+\sum_{j=1}^{i-1}m_{c,j} & *=c
\end{array}\right.
\end{displaymath}
Hence
\begin{equation}
B^{T}X_{*,i}=(T^{-1}B)^{H}\left[\begin{array}{c} 0_{M_{*,i}-1}\times p_{*}(i) \\ Z_{*,i} \\
0_{(n-M_{*,i})\times p_{*}(i)} \end{array}\right]
\label{eqn:18}
\end{equation}

Assume now that the input matrix $B$ is given by Equation (\ref{eqn:4}). Then,
\begin{eqnarray}
B^{T}X_{*,i}&=&{\rm\bf col}^{H}\left\{ {\rm\bf col}\left\{\hat{B}_{r,i}|_{i=1}^{k_{r}}\right\},\;\;
{\rm\bf col}\left\{\hat{B}_{c,i}|_{i=1}^{k_{c}}\right\}\right\}\times\nonumber\\
& & \left[\begin{array}{c} 0_{M_{*,i}-1}\times p_{*}(i) \\ Z_{*,i} \\
0_{(n-M_{*,i})\times p_{*}(i)} \end{array}\right]\nonumber\\
&=&\hat{B}_{*,i}Z_{*,i} \nonumber\\
&=& \tilde{B}_{*,i}
\label{eqn:19}
\end{eqnarray}
Therefore, when the matrix $\tilde{B}_{*,i}$ is of FCR for each $*=r,\;c$ and each $i=1,2,\cdots,k_{*}$, it can be declared from Lemma 2 that the system is controllable.

On the contrary, assume that the system is controllable. Then, according to Lemma 2, the matrix $B^{T}X_{*,i}$ is necessarily of FCR for each feasible $*$ and $i$. Construct matrices $\hat{B}_{*.i}$, $*=r,c$, $i=1,2,\cdots,k_{*}$, such that the following equality is satisfied
\begin{equation}
{\rm\bf col}\left\{ {\rm\bf col}\left\{\hat{B}_{r,i}|_{i=1}^{k_{r}}\right\},\;\;
{\rm\bf col}\left\{\hat{B}_{c,i}|_{i=1}^{k_{c}}\right\}\right\}=T^{-1}B
\label{eqn:20}
\end{equation}
This construction is obviously always possible. Moreover, from Equation (\ref{eqn:18}), the corresponding $\tilde{B}_{*,i}$ is always of FCR.
This completes the proof. \hspace{\fill}$\Diamond$

From the definition of the matrix $B$ in Equation (\ref{eqn:4}), it is clear that it has just $p_{max}$ columns. Hence, the results of Theorem 2 in fact give a complete parametrization for all the input matrices $B$ that has the minimal number of inputs and construct a controllable system with the STM $A$. On the other hand, the requirement of guaranteing that the resulting matrix $B$ is real can be simply achieved by setting each $\hat{B}_{r,i}$ to be real, $i=1,2,\cdots,k_{r}$, and setting each $\hat{B}_{c,i+k_{c}/2}$ as $\hat{B}_{c,i+k_{c}/2}=\bar{\hat{B}}_{c,i}$ for $i=1,2,\cdots,k_{c}/2$.

From the proof of Theorem 2, it is obvious that its conclusions are valid for an arbitrary system that has inputs more than $p_{max}$. It can therefore be declared that when $p_{max}$ is replaced by an integer $q$ not smaller than $p_{max}$, Equation (\ref{eqn:4}) also gives a complete parametrization for all the input matrices $B$ that have $q$ columns and  construct a controllable system with the STM $A$.

For a prescribed STM $A$, similar results can be obtained for parameterizing all the output matrices of an observable system with its output number not smaller than $p_{max}$. These results can be simply obtained through the duality between system controllability and observability, and the details are therefore omitted.

\section{Concluding Remarks}

In this paper, we have investigated the problem about the minimal number of inputs/outputs under the requirements that there exists an input/output matrix, such that the associated linear time invariant system is controllable/observable. It has been made clear that this number is equal to the maximum geometric multiplicity of the state transition matrix, which can be calculated in principle.

These conclusions are in sharp contrast to the minimal controllability/observability problems attacked in \cite{pka16,scl15}, which have been proved to be NP-hard, and even difficult to be approximately solved within a multiplicative factor. These results suggest that as long as controllability/observability is concerned, it is preferable to construct a system with its state transition matrix having distinct eigenvalues, in the sense of hardware cost reduction. In addition, a parametrization is given for all the input/output matrices that has the minimal inputs/outputs and construct a controllable/observable system with a prescribed STM.

As a further topic, it is interesting to see whether or not these results can be extended to situations in which there are some other structure restrictions on the system input/output matrix, which is often met in practical applications \cite{pka16,wj01}, as well as to situations in which subsystems are connected through their outputs like those discussed in \cite{zhou15,zz16}. The latter is thought to be a more natural way in describing dynamics of a large scale system.

\hspace*{-0.45cm}{\bf Acknowledgements.} Discussions with Prof.Basar of the University of Illinois at Urbana and Champaign, and Prof.Antsaklis of the University of Notre Dame, are greatly appreciated.

\end{document}